\documentclass{caosp306}

\usepackage{graphicx}

\usepackage{natbib}
\pubyear{2019}
\volume{49}
\volnumber{1}
\firstpage{7}
\received{Sep 26, 2018}
\accepted{November 15, 2018}

\def\BibTeX{{\rm B\kern-.05em{\sc i\kern-.025em b}\kern-.08em
             T\kern-.1667em\lower.7ex\hbox{E}\kern-.125emX}}

\begin{document}

%
\htitle{Estimating the black hole mass of NGC 1313 X-1}
\hauthor{C.-Y.\,Huang}

\title{Estimating the black hole mass of NGC 1313 X-1 based on the spectral and timing properties}


%
\author{
        C.-Y.\,Huang \inst{1}
       }

%
\institute{
           School of Physics and Optoelectronic Engineering, Yangtze University, Jingzhou 434023, China \email{hcy@yangtzeu.edu.cn}
          }


\maketitle

\begin{abstract}
The discovery of 3:2 quasi-periodic oscillations (QPOs) in the ultraluminous X-ray source NGC 1313 X-1 suggests it harbors an intermediate-mass black hole (IMBH). We test this numerically by modelling the 3:2 QPOs and the associated X-ray spectrum based on the epicyclic resonance model and a disk-corona model with large-scale magnetic fields generated by the Cosmic Battery mechanism. The combined QPO-frequency and spectral fitting indicates that the BH mass ranges  from 2524$M_{\rm \odot}$ to 6811$M_\odot$ confirming its IMBH nature, and the BH spin is probably higher than $\sim$ 0.3.
\keywords{accretion, accretion disks -- black hole physics -- magnetic fields -- stars: individual:
NGC 1313 X-1}
\end{abstract}

%
\section{Introduction}
\label{intr}
Ultraluminous X-ray sources (ULXs) are non-nuclear point-like X-ray sources in nearby galaxies with luminosities greater than $ \sim10^{39} {\rm erg}\ {\rm s}^{-1}$. The nature of the compact objects in ULXs is still under debate. They are generally thought to be powered by the intermediate-mass ($10^2$ -- $10^4 M_\odot$) black holes (IMBHs) since the luminosity exceeds the Eddington limit for a typical stellar-mass ($\sim10 M_\odot$) black hole (BH), where $M_\odot$ is the solar mass. However, the spectral analyses based on high quality observations indicate that they may harbor stellar-mass BHs accreting at near-Eddington or super-Eddington rate \citep[e.g.][]{2006MNRAS.368..397S,2009MNRAS.397.1836G,2017ApJ...849..121J}. The super-Eddington phenomenon could also be explained by the beaming geometry or the relativistic beaming of jet emission \citep[e.g.][]{2001ApJ...552L.109K,2007MNRAS.377.1187P,2002A&A...382L..13K,2006MNRAS.372..630F,2007IAUS..238..247K}. Recently, the discovery of coherent X-ray pulsations from some ULXs indicates that the central engines are accreting neutron stars \citep{2014Natur.514..202B,2017MNRAS.466L..48I,2016ApJ...831L..14F,2017Sci...355..817I,2017A&A...608A..47K}. See \cite{2017ARA&A..55..303K} for a review. A direct mass measurement is needed to clarify the question of what are the compact objects harbored in ULXs.

Timing properties provide us an important clue to constrain the mass and spin of the compact objects. The discovery of 3:2 quasi-periodic oscillation (QPO) pairs in two ULXs indicates that they harbor IMBHs. \citet{2014Natur.513...74P} reported a 3:2 QPO pair at frequencies 5 and 3.3 Hz detected in ULX M82 X-1 and estimated the mass of the compact object to be $428\pm105M_\odot$ using the inverse correlation between the BH mass and the frequencies of high-frequency QPOs (HFQPOs) observed in galactic BH X-ray binaries (BHXBs). Soon afterwards \citet{2015ApJ...811L..11P} detected another 3:2 QPO pair at frequencies 0.45 and 0.3 Hz in \emph{XMM-Newton} observations of ULX NGC 1313 X-1. They found that the power density spectrum of NGC 1313 X-1 is analogous to that of BHXBs in the steep power-law (SPL) state and estimated the BH mass to be $5000 M_\odot$ using the inverse mass-to-HFQPO frequency scaling.

\cite{2018MNRAS.473..136J} applied the X-ray scaling method to measure the BH mass in ULXs with multiple X-ray observations and their estimated BH mass of NGC 1313 X-1 was in the range 295 -- 6166 $M_{\rm \odot}$.

Both the QPO scaling and the X-ray scaling methods imply NGC 1313 X-1 harbors an IMBH. However, quantitative and numerical analyses are needed to test this result. In this paper, we estimate the BH mass in NGC 1313 X-1 by modelling the 3:2 QPOs and the associated X-ray spectrum simultaneously based on the epicyclic resonance model (ERM) and a magnetic disk-corona model developed by \cite{2016MNRAS.457.3859H}.

The paper is organised as follows. In section 2, we give a description of the model. In section 3, the 3:2 QPO frequencies and the associated X-ray spectrum are fitted and the range of the BH mass is obtained. We discuss and summarize in section 4. The geometric units $G=c=1$ are used throughout this paper.

\section{Model description}

\subsection{The magnetic disk-corona model}
The energy transferred from the rotating BH and the plunging region to the inner disk is considered through the magnetic connection (MC) process by the large-scale magnetic fields generated by the Poynting-Robertson Cosmic Battery (PRCB) mechanism \citep{1998ApJ...508..859C,2006ApJ...652.1451C,2012A&A...538A...5K}. In the PRCB scenario, electrons in the inner accretion disk feel a much stronger radiation-drag force than the ions due to a much larger Thompson cross-section leading to an electric current loop generated near the inner edge of the accretion disk. The current intensity can be estimated as \citep{2016MNRAS.457.3859H}
%
%
\begin{equation}
 \label{eq1}
I=en\Delta t \Delta S F\sigma_{\rm T}v_{\rm k}/m_{\rm e},
\end{equation}
where $\Delta t$ and $\Delta S$ are the average scattering time of electrons with photons and the cross-section of the current, respectively, $F$ is the radiation flux from the relativistic accretion disk \citep{1973blho.conf..343N,1974ApJ...191..499P}, $n$ is the number density of electrons, $v_{\rm k}$ is the Kepler velocity, $\sigma_{\rm T}$ is the Thompson cross-section and $e$ is the electric quality of an electron. Following \cite{2016MNRAS.457.3859H}, we set $\Delta t=10^{-9}{\rm s}$ and $\Delta S=(0.01M_{\rm BH})^2$, where $M_{\rm BH}$ is the BH mass. The values of $F$, $n$ and $v_{\rm k}$ are all evaluated at the radius $r_{\rm PRCB}$ where the current is generated. As suggested by \cite{2012A&A...538A...5K}, the current is generated near the innermost stable circular orbit (ISCO) of the disk. Fitting the 3:2 HFQPO pairs and the associated spectrum from BHXBs by \cite{2016MNRAS.457.3859H} also indicates that the current is very near the ISCO. For simplicity, We set $r_{\rm PRCB}=1.01 r_{\rm ISCO}$ in this work, where $r_{\rm ISCO}$ is the radius of ISCO.

The magnetic fields generated by the PRCB current can be calculated in the frame of general relativity \citep{1979JPhA...12..839L,2002PhRvD..65h4047L}. As shown in \cite{2016MNRAS.457.3859H}, two types of MC configurations are created by the PRCB current with one connecting the BH horizon and the disk and the other connecting the plunging region and the inner disk. The rotational energy of both the BH and the plunging region can be transferred to the inner disk by these two MC processes. The energy dissipated at unit disk area caused by MC is expressed as \citep{2002ApJ...567..463L}
\begin{equation}\label{eq2}
  Q_{\rm MC}^+=r^{-1}S^{-2}(-\frac{{\rm d}\Omega_{\rm D}}{{\rm d}r})\int^r_{r_{\rm in}}SH_{\rm MC}r{\rm d}r,
\end{equation}
where $S=E^\dag-\Omega_{\rm D}L^\dag$ with $E^\dag$, $L^\dag$ and $\Omega_{\rm D}$ are, respectively, the specific energy, the specific angular momentum and angular velocity of a test particle in the accretion disk \citep{1972ApJ...178..347B}. In equation~(\ref{eq2}), $H_{\rm MC}\equiv\frac{1}{4\pi r}\frac{{\rm d}T}{{\rm d}r}$ is the flux of angular momentum transferred to the disk, where $T$ is the torque exerting on the disk surface by MC \citep{2002ApJ...567..463L}, and $r_{\rm in}$ is the inner most disk radius receiving energy from the plunging region.

The gravitational power dissipated in unit disk surface is \citep{1974ApJ...191..499P}
\begin{equation}\label{eq3}
    Q_{\rm DA}^+=\frac{\dot{M}}{4\pi r}f,
\end{equation}
where $\dot{M}$ is the mass accretion rate of the disk and $f$ is a function of $a_*$, $r$ and $M_{\rm BH}$ as given in \cite{1974ApJ...191..499P}, where $a_*\equiv J/M_{\rm BH}^2$ is the dimensionless BH spin defined in terms of $M_{\rm BH}$ and BH angular momentum $J$.

The corona is assumed to be heated by the reconnection of the small-scale magnetic field that generated by the buoyancy instability in the disk \citep{1998MNRAS.299L..15D,2002ApJ...572L.173L}. The power dissipated in the corona is \citep{2002ApJ...572L.173L}
\begin{equation}\label{eq4}
     Q_{\rm cor}^+=\frac{B_{\rm d}^2}{4\pi}v_{\rm A},
\end{equation}
where $B_{\rm d}$ is the small-scale magnetic field in the disk, and $v_{\rm A}\equiv B_{\rm d}/\sqrt{4\pi\rho}$ is the Alfv$\acute{\rm e}$n speed, where $\rho$ is the mass density of the disk.

Part of the energy dissipated in the disk is radiated as black body, and the other part is released into the corona. Thus, the energy equation for the disk is
\begin{equation}\label{eq5}
    Q_{\rm DA}^++Q_{\rm MC}^+-Q_{\rm cor}^+=\frac{4\sigma T_{\rm d}^4}{3\tau_{\rm d}},
\end{equation}
where $T_{\rm d}$ is the temperature in the mid-plane of the disk, $\sigma$ is the Stefan-Boltzmann constant, and $\tau_{\rm d}$ is the optical depth in the vertical direction of the disk which is related to the mean opacity $\kappa$ as
\begin{equation}\label{eq6}
\tau_{\rm d}=\rho h \kappa=\rho h (\kappa_{\rm ff}+\kappa_{\rm es}),
\end{equation}
where $h$ is the half-thickness of the disk, $\kappa_{\rm ff}$ and $\kappa_{\rm es}$ are the contributions of free-free transitions and electron scattering respectively, and  $\kappa_{\rm ff}=6.4\times10^{22}\rho T_{\rm d}^{-7/2} \,{\rm cm}^2 \,{\rm g}^{-1}$ and $\kappa_{\rm es}=0.4\, {\rm cm}^2 \,{\rm g}^{-1}$ are adopted \citep{1973blho.conf..343N}.  

The equation for the vertical pressure balance of the disk is \citep{1973blho.conf..343N}
\begin{equation}\label{eq7}
h=P^{1/2}\rho^{-1/2}r^{3/2}M_{\rm BH}^{-1/2}\mathcal{A}\mathcal{B}^{-1}\mathcal{C}^{1/2}\mathcal{D}^{-1/2}\mathcal{E}^{-1/2},
\end{equation}
where $P=P_{\rm gas}+P_{\rm rad}+P_{\rm mag}$, and $P_{\rm gas}$, $P_{\rm rad}$ and $P_{\rm mag}$ are,  respectively, the gas pressure, radiation pressure and magnetic pressure of the disk. $\mathcal{A}$, $\mathcal{B}$, $\mathcal{C}$, $\mathcal{D}$ and $\mathcal{E}$ are the relativistic correction factors in Kerr metric \citep{1973blho.conf..343N}.

Magnetohydrodynamics (MHD) simulations reveal that the viscosity in the accretion disk is related to the tangled small-scale magnetic fields. However, the detailed physics of the generation of the small-scale magnetic fields is still unclear. The famous `$\alpha$-prescription' proposed by \cite{1973A&A....24..337S} is widely used in various theoretical works. Observations favor the interior viscous stress $t_{r\varphi}$ proportional to the gas pressure \citep{2004ApJ...607L.107W}. Therefore we adopt this $\alpha$-prescription in our calculation:
\begin{equation}\label{eq8}
    t_{r\varphi}=P_{\rm mag}=\frac{B_{\rm d}^2}{8\pi}=\alpha P_{\rm gas},
\end{equation}
where $\alpha$ is the viscosity parameter.

The continuity equation of the disk is \citep{1973blho.conf..343N}
\begin{equation}\label{eq9}
    \dot{M}=-4\pi rh\rho v_r \mathcal{D}^{1/2},
\end{equation}
where $v_r$ is the radial velocity of the accretion flow at radius $r$.
Incorporating the energy and angular-momentum equations for the disk, the stress $t_{r\varphi}$ is related to the dissipated energy as follow \citep{1974ApJ...191..499P,2002ApJ...567..463L}
\begin{equation}\label{eq10}
   r h t_{r\varphi}=S(-\frac{{\rm d}\Omega_{\rm D}}{{\rm d}r})^{-1}(Q_{\rm DA}^++Q_{\rm MC}^+).
\end{equation}

The equation of state for the gas in the disk is
\begin{equation}\label{eq11}
    P_{\rm rad}+P_{\rm gas}=\frac{1}{3}a_0T_{\rm d}^4+\frac{\rho k T_{\rm d}}{\mu m_{\rm p}},
\end{equation}
where $a_0$ is the radiation constant, $k$ is the Boltzmann constant, $m_{\rm p}$ is the proton mass, and $\mu=0.615$ is adopted being the mean atomic mass.

The electrons in the corona are heated by the magnetic reconnection and are cooled via the inverse Compton scattering by the soft photons from the cold disk. Thus, the energy equation of the corona is \citep{2002ApJ...572L.173L}
\begin{equation}\label{eq12}
    Q_{\rm cor}^+=Q_{\rm comp}^-=\frac{4kT_{\rm e}}{m_{\rm e}}\tau U_{\rm rad},
\end{equation}
where $U_{\rm rad}=a_0 T_{\rm d}^4$ is the energy density of the soft photon field, $T_{\rm e}$ is the temperature of the corona, $m_{\rm e}$ is the electron mass, and $\tau$ is the optical depth of the corona.

Following \cite{2016MNRAS.457.3859H}, we take the optical depth, the height and the outer boundary of the corona as $\tau=1$, $l = 20 M_{\rm BH}$ and $r_{\rm out}=100 M_{\rm BH}$, respectively. Solving equations (\ref{eq1}) -- (\ref{eq12}) numerically, we can obtain the global solutions of the disk-corona system with MC process. The X-ray spectrum of the system can be simulated using Monte Carlo method \citep{2009MNRAS.394.2310G}.

\subsection{The 3:2 QPO model}

There is no consensus on the mechanism of QPO generation although a variety of models have been proposed \citep[see][for a review]{2014SSRv..183...43B}. We interpret 3:2 QPO pairs based on ERM \citep{2001A&A...374L..19A} since it naturally explains the small integer ratios of 3:2 QPO frequencies.

In the general relativistic frame, a test particle in the equatorial plane around a compact object has three basic oscillation frequencies, namely, the orbital frequency $\nu_\phi$, the vertical epicyclic frequency $\nu_\theta$ and the radial epicyclic frequency $\nu_r$, which are functions of $a_*$, $r$ and $M_{\rm BH}$ given as follows \citep{1972ApJ...178..347B,1987PASJ...39..457O,1990PASJ...42...99K,2006csxs.book...39V}
\begin{equation}\label{eq13}
  \nu_\phi=[2\pi M_{\rm BH}(\tilde{r}^{3/2}+a_*)]^{-1},
\end{equation}
\begin{equation}\label{eq14}
  \nu_\theta=\nu_\phi(1-4a_*\tilde{r}^{-3/2}+3a^2_*\tilde{r}^{-2})^{1/2},
\end{equation}
\begin{equation}\label{eq15}
  \nu_r=\nu_\phi(1-6\tilde{r}^{-1}+8a_*\tilde{r}^{-3/2}-3a^2_*\tilde{r}^{-2})^{1/2},
\end{equation}
where $\tilde{r}\equiv r/M_{\rm BH}$.

In ERM, resonances occur at specific radii of the disk if the vertical and the radial epicyclic frequencies have small integer ratios resulting in oscillations of the plasmoid. As argued by \cite{2010MNRAS.403.1978H,2016MNRAS.457.3859H}, the severe damping forces on the oscillations can be overcome by transferring energy from the spinning BH to the inner disk via the MC process therefore the resonances persist and emit X-rays with sufficient amplitude and coherence to produce the observed QPOs.

Applying to NGC 1313 X-1, we treat the frequency of the 0.45 Hz QPO as $\nu_\theta$ and the 0.3 Hz as $\nu_r$.
For a given BH spin, the BH mass and the resonance radius $\tilde{r}_{\rm res}$ can be obtained by solving the equations
\begin{equation}\label{eq16}
  \left\{
    \begin{array}{l}
      \nu_\theta/\nu_r=3/2,  \\
      \nu_\theta=0.45.
    \end{array}
  \right.
\end{equation}

\section{Fitting the 3:2 QPO frequencies and the associated X-ray spectrum}

The 0.45 and 0.3 Hz QPOs of NGC 1313 X-1 were detected simultaneously on 2012 December 16 in the \emph{XMM-Newton} observation (obsID: 0693850501) at energy band of 0.3 -- 10 keV \citep{2015ApJ...811L..11P}. We first fit the 3:2 QPO frequencies and then fit the 0.3 -- 10 keV spectrum from this \emph{XMM-Newton} observation.

In the spectral simulation, the distance to the source $d=4.61{\rm Mpc}$ \citep{2015ApJ...799...19Q} is adopted and the inclination $i$ of the system is assumed to be $70^\circ$ for an analogy with the X-ray binary systems since almost all the BHXBs displaying HFQPOs have high inclinations \citep{2015PhR...548....1M}, which is close to the values (66$^\circ$ -- 76$^\circ$) obtained by spectral fitting models of \cite{2013ApJ...778..163B}.

Since the BH spin of NGC 1313 X-1 has not been measured, we consider two extremes, the Schwarzschild BH with $a_*=0$ and the extreme Kerr BH with $a_*=0.998$. The corresponding lower and upper limits of the BH mass obtained by solving equation~(\ref{eq16}) are $2024M_\odot$ and $6811M_\odot$, respectively. Based on the values of $m_{\rm BH}$ ($m_{\rm BH}\equiv M_{\rm BH}/M_\odot$) and $a_*$, we then fit the spectrum by adjusting the parameters $\dot{m}$, $\alpha$ and $n_{\rm H}$, where $\dot{m}\equiv\dot{M}/\dot{M}_{\rm Edd}$ and $\dot{M}_{\rm Edd}\simeq1.4\times10^{18} m_{\rm BH} \ {\rm g} \ {\rm s}^{-1}$ is the Eddington accretion rate, and $n_{\rm H}$ is the hydrogen column density.

\begin{table}[t]
\small
\begin{center}
\caption{Parameters for fitting the 3:2 QPO frequencies and the associated X-ray spectrum.}
\label{t1}
\begin{tabular}{llllll}
\hline\hline
     $a_*$   & $m_{\rm BH}$ & $\tilde{r}_{\rm res}$ & $\dot{m}$ & $\alpha$ & $n_{\rm H}$($10^{22} {\rm cm}^{-2}$) \\
\hline
0.3    & 2524 & 9.13 & 0.38    & 0.39 & 0.45    \\
0.998  & 6811 & 3.89 & 0.00032 & 0.22 & 0.28    \\
\hline\hline
\end{tabular}
\end{center}
\end{table}

\begin{figure}[t]
\centerline{\includegraphics[width=0.6\textwidth,clip=]{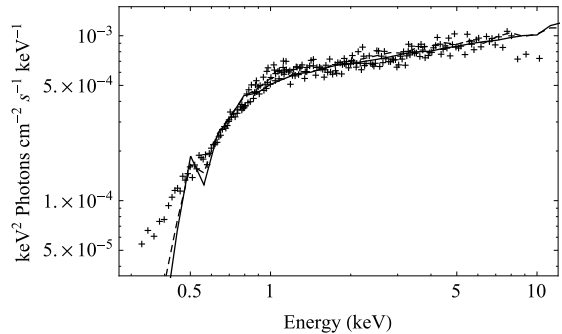}}
\caption{The simulated and observed spectra of NGC 1313 X-1 associated with the 3:2 QPOs. The solid and dashed lines represent the model spectra with $a_*=0.3$ and $a_*=0.998$, respectively. The symbol ``+" denotes the observed spectrum (without error bars) with data taken from \cite{2013ApJ...778..163B}.}
\label{f1}
\end{figure}

It turns out that a larger BH mass leads to a better spectral fitting and the spectrum can not be fitted well if $a_*\la0.3$ since the BH mass is too small. The parameters corresponding to the lower and upper limits of the values of $a_*$ and $m_{\rm BH}$ for fitting the 3:2 QPO frequencies and the associated X-ray spectrum are listed in Table\,\ref{t1}. The corresponding simulated spectra are plotted in Fig.\,\ref{f1}, which fit very well into the observed one in the 0.5 -- 10 keV band. The deviation below 0.5 keV may be due to that we do not consider the synchrotron and bremsstrahlung radiation from the corona in our model.

The simulated spectrum is more sensitive to $\dot{m}$ for a larger $a_*$ due to the stronger MC effect by the large-scale magnetic fields generated by PRCB current which is proportional to $\dot{m}$ as indicated by equation~(\ref{eq3}).

Inspecting Table~\ref{t1}, one can find that a larger BH mass is needed to fit the 3:2 QPO frequencies for a higher BH spin. And larger values of $\dot{m}$ and $\alpha$ are needed to fit the spectrum for a lower BH spin and a smaller BH mass, which means both the soft component, which is proportional to $\dot{m}$ as indicated by equation~(\ref{eq3}), and the hard component, which is proportional to $\alpha$ as indicated by equation~(\ref{eq8}) and equation~(\ref{eq4}), should be enhanced to fit the observed spectrum.

\section{Summary and discussion}

In this paper, we modelled the 0.45 and 0.3 Hz QPOs and the associated 0.3 -- 10 keV X-ray spectrum of NGC 1313 X-1 based on ERM and a disk-corona model with PRCB. The combined QPO-frequency and spectral fitting indicates that the BH mass lies in the range 2524 -- 6811$M_\odot$ and the BH spin is probably higher than $\sim0.3$. This is the first time to model the combined spectral and timing properties of NGC 1313 X-1 numerically and self-consistently. The result confirms that NGC 1313 X-1 contains an IMBH consistent with the results of \cite{2015ApJ...811L..11P} and \cite{2018MNRAS.473..136J}.

\begin{figure}[t]
\centerline{\includegraphics[width=0.6\textwidth,clip=]{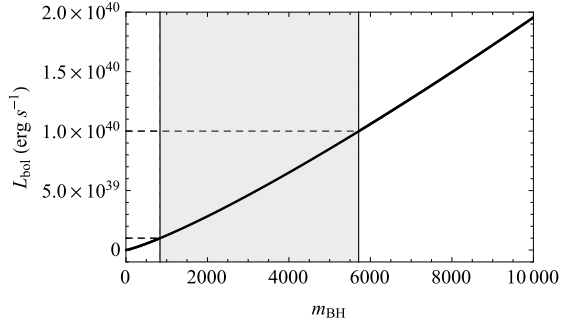}}
\caption{The curve of the bolometric luminosity as a function of BH mass for the PRCB magnetic fields at ISCO become saturated. The shaded area indicates the BH mass ranges between 839$M_\odot$ and 5719$M_\odot$ for the bolometric luminosity ranges from $10^{39}$ to $10^{40} \ {\rm erg\ s}^{-1}$.}
\label{f2}
\end{figure}

We find that the magnetic fields play an important role in the spectral fitting. The simulated spectra are in accordance with the observed one only when the intensity of the magnetic fields at $r_{\rm ISCO}$ generated by PRCB is comparable to that at the BH horizon evaluated by the balance of the ram pressure and the magnetic pressure. This means that the magnetic fields become saturated and the disk is in the magnetically arrested disk (MAD) state \citep{2003ApJ...592.1042I}. Our model suggests that ULXs especially those brightest, like NGC 1313 X-1, may be close to or reach the MAD state. If this is indeed the case, we can predict the bolometric luminosity of an ULX by figuring out the accretion rate to make the magnetic fields become saturated for a given BH mass, or, conversely, the BH mass can be obtained with the saturated magnetic fields for a given  bolometric luminosity. 

Let's make a simple estimate. For $a_*=0.9$ and $\alpha=0.3$, the intensity of the PRCB magnetic fields $B_{\rm PRCB}$ at $r_{\rm ISCO}$ in our model can be expressed as 
\begin{equation}\label{eq17}
 B_{\rm PRCB}=4.5\times10^7\left(\frac{\dot{m}}{0.1}\right)^{1.5}\left(\frac{m_{\rm BH}}{5000}\right)^{-0.7} {\rm Gauss}.
\end{equation}
The MAD magnetic fields can be estimated by the balance of the ram pressure of the innermost part of the accretion flow and the magnetic pressure on the BH horizon \citep{1997rja..proc..110M,2002MNRAS.335..655W,2009MNRAS.394.2310G}:
\begin{equation}\label{eq18}
 B_{\rm MAD}=\sqrt{2\dot{M}/r_{\rm H}}=6.1\times10^6\left(\frac{\dot{m}}{0.1}\right)^{0.5}\left(\frac{m_{\rm BH}}{5000}\right)^{-0.5} {\rm Gauss},
\end{equation}
where $r_{\rm H}=M_{\rm BH}(1+\sqrt{1-a^2_*})$ is the radius of the BH horizon. Assuming the radiation efficiency $\eta=0.1$ and  the bolometric luminosity $L_{\rm bol}=\eta \dot{M}$, let $B_{\rm PRCB}=B_{\rm MAD}$, we have
\begin{equation}\label{eq19}
 L_{\rm bol}=8.5\times10^{39}\left(\frac{m_{\rm BH}}{5000}\right)^{1.2} {\rm erg}\ {\rm s}^{-1}.
\end{equation}
If $L_{\rm bol}$ is between $10^{39}$ and $10^{40}$ erg s$^{-1}$, then the constrained BH mass is between 839$M_\odot$ and 5719$M_\odot$, as indicated by the grey shaded area in Fig.~\ref{f2}.

In the spectral fitting, we only considered the 0.3 -- 10 keV \emph{XMM-Newton} spectrum associated with the 3:2 QPOs. The \emph{NuSTAR} spectrum shows a clear cutoff above 10 keV and decays up to $\sim$ 30 keV, which can be described by a phenomenological model that includes a multi-color disk plus a two-component corona composed of a cold, optically thick medium and a second, hot and optically thin one~\citep{2013ApJ...778..163B}. We will improve our model by considering multi-temperature corona with different geometries (e.g. spherical corona, ``lamp-post" corona) in future works to understand both the \emph{XMM-Newton} and \emph{NuSTAR} spectra and to obtain more accurate and reliable measurements for the BH mass and spin of NGC 1313 X-1 and other BH systems.

\acknowledgements
This work has been supported by the National Natural Science Foundation of China (Grant No.11403003). We thank the anonymous referee for his (her) helpful suggestion.

\bibliography{demo_caosp306}

%

\end{document}